# Whale Optimization Algorithms based fractional order fuzzy PID controller for Depth of Anesthesia

Amin Behboudifar, Chen Jing

*Abstract*— One of the most important surgical factors is Depth of Anesthesia (DOA) control in patients. The main problem is to overcome the uncertainty and nonlinearity of the system, due to different physiological parameters of the patient's body and maintain DOA of patients in desired range during surgery. This study demonstrates a fractional order fuzzy PID controller (FOFPID) and fractional order PID controller (FOPID) to the problem. The Whale Optimization Algorithms (WOA) is used to optimized the parameters of proposed controllers. The orders of derivative and integral fractional controller is achieved by WOA. The results indicate that FOFPID has a better performance than FOPID. To check the performance of the controllers in presence of uncertainty, physiological logical model of 8 patients has been investigated. The modeling is based on Pharmacodynamic and Pharmacokinetic model. The results show the performance of the proposed method.

## I. Introduction

Anesthesia is a state that patients in this state are unconsciousness caused by drug infusion not to feel pain during surgery. One of the problems that may sometimes occur during surgery, is consciousness during surgery. In addition to the problem, over-administration of anesthetic drugs can cause problems such as delaying anesthesia and poisoning, or, in the worst case, causing patient death. Also, one of the main problems of anesthesiologists is uncertainty about the DOA and therefore it is necessary to use methods that help in DOA control [1].

Precise control of the Depth of Anesthesia (DOA) is paramount in surgical settings to ensure patient unconsciousness and pain suppression while minimizing the risk of adverse outcomes [2]. This challenge lies at the intersection of control engineering, artificial intelligence (AI), and clinical pharmacology—fields that collectively impact patient safety, healthcare efficiency, and the development of medical technologies. The advent of AI-driven closed-loop control systems has revolutionized anesthesiology by enabling real-time, adaptive drug administration that accounts for diverse physiological responses among patients. These advancements are particularly significant for researchers, clinicians, and medical device innovators aiming to enhance surgical precision and reduce healthcare costs [3].

The primary challenge in DOA control is managing the inherent nonlinearity and uncertainty of human physiological systems, which is largely driven by inter-patient variability. The Bispectral Index (BIS), a key indicator of consciousness, must be maintained within a safe range of 40–60 (optimal at 50) to ensure effective anesthesia without risking intraoperative awareness or drug overdose. Conventional approaches, such as manual or open-loop drug administration, often fail to adapt to patient-specific dynamics or surgical disturbances—thereby compromising safety and recovery. The demand for robust, automated control solutions is growing in high-stakes surgical environments, aligning with the increasing emphasis on intelligent systems in biomedical engineering [4].

Previous research in DOA control has explored a range of methodologies to address the complexities of anesthetic drug delivery [5]. Early open-loop systems, lacking real-time feedback, were limited in managing patient variability. The introduction of closed-loop control brought significant improvements, with techniques such as PID control, neural network-based adaptive control, and sliding-mode control showing better BIS regulation. Fuzzy logic, a widely used AI method, has gained popularity for its ability to model uncertainty through linguistic rules and membership functions, effectively mimicking human decision-making. Methods of increasing the domain of attraction of nonlinear systems can also be used to enhance the performance of control systems in the presence of model uncertainties and sensory noises. Methods of increasing the domain of attraction of nonlinear systems [6] can also be used to enhance the performance of control systems in the presence of model uncertainties and sensory noises [7]. Additionally, fractional-order control has emerged as a promising alternative, offering greater flexibility via non-integer derivatives and integrals, which enhance performance beyond traditional controllers.

Recent advancements have increasingly incorporated AI techniques to optimize DOA control systems. Bio-inspired algorithms such as particle swarm optimization and genetic algorithms have been employed to fine-tune controller parameters. The Whale Optimization Algorithm (WOA)—an AI-driven approach inspired by the hunting behavior of humpback whales—has proven effective in optimizing complex, high-dimensional parameter spaces [8]. Moreover, combining fuzzy logic with fractional-order control—as in Fractional Order Fuzzy PID (FOFPID) controllers—leverages both AI's adaptability and the precision of fractional calculus to manage nonlinear systems. However, the application of AI-optimized FOFPID controllers for DOA, particularly under varied patient conditions, remains relatively underexplored [9].

Despite these advancements, current DOA control strategies often struggle to deliver robust performance across a wide range of patient profiles due to limitations in handling system nonlinearity and physiological uncertainty. Many existing methods rely on integer-order controllers or heuristic tuning processes, which may not fully capture the complex dynamics of anesthesia delivery. While fuzzy and fractional-order controllers offer theoretical advantages, their real-world implementation is hindered by the difficulty of precise parameter tuning and the need to adapt to patient-specific variability. This gap limits the clinical adoption of advanced controllers, where both reliability and generalizability are essential. This study is motivated by the need to develop a robust, AI-driven control solution to address these challenges and improve DOA precision [10].

This research proposes a Fractional Order Fuzzy PID (FOFPID) controller, optimized using the Whale Optimization Algorithm (WOA), to achieve precise and robust DOA control across multiple patient models. The primary objectives are to maintain the BIS within the 40–60 range and to evaluate controller performance under physiological uncertainties. The key contributions of this work include: (1) an AI-driven FOFPID controller that combines the adaptability of fuzzy logic with the precision of fractional-order dynamics, (2) WOA-based optimization of controller parameters, including fractional orders and fuzzy membership functions, and (3) a comprehensive evaluation against a Fractional Order PID (FOPID) baseline using pharmacokinetic and pharmacodynamic models for eight distinct patients. These contributions aim to advance automated anesthesia delivery, offering a scalable, AI-enhanced solution suitable for clinical practice.

The remainder of this paper is structured as follows: Section 2 presents the system model, Section 3 describes the FOFPID and WOA methodology, Section 4 discusses simulation results, and Section 5 concludes with key findings and outlines future directions.

## II. LITERATURE SURVEY

The control of Depth of Anesthesia (DOA) has emerged as a pivotal research domain, integrating biomedical engineering, control theory, and pharmacology to ensure safe and effective anesthetic delivery during surgical procedures. Early strategies relied on open-loop systems, in which anesthesiologists manually adjusted drug infusion rates based on clinical observations. These methods often resulted in inconsistent outcomes due to patient-specific physiological variations. The advent of closed-loop control systems marked a significant advancement by incorporating real-time feedback mechanisms—such as the Bispectral Index (BIS)—to dynamically regulate anesthesia levels. Foundational studies established pharmacokinetic and pharmacodynamic models that simulate patient responses to anesthetics like Propofol, laying a robust framework for modern control methodologies. These foundational efforts enabled the integration of advanced computational techniques, including artificial intelligence (AI), to address the nonlinearity and uncertainty inherent in DOA systems.

[11] This paper optimizes 2D steel chevron-braced frames for seismic performance using an ant colony optimization algorithm and nonlinear pushover analysis, demonstrating improved energy dissipation and load-carrying efficiency compared to traditional designs. [12] This paper combines performance-based pushover analysis with the Artificial Bee Colony (ABC) optimization algorithm to efficiently optimize seismic design of 2D reinforced concrete wall-frames, focusing on weight reduction while considering nonlinear behavior and reinforcement arrangements

Ahmadi et al. survey recent advances in unsupervised time-series analysis using autoencoders and vision transformers, detailing key architectures and diverse real-world applications [13]. This paper presents a sensor-based simulation approach for predicting global lateral displacement in reinforced concrete bridge columns during seismic events using flexural and bar–slip models [14]. This paper presents a framework that uses smart IoT dashcams and a digital twin to remotely monitor and detect damaged road furniture, enabling real-time updates and proactive maintenance to enhance road safety [15]. This paper introduces a probabilistic inference-based framework using a sequential ensemble Kalman smoother to achieve optimal control of convolutional neural networks for high-dimensional dynamic systems [16]. This paper proposes a hyperdimensional computing-based framework for IoT intrusion detection, achieving 99.54% accuracy on the NSL-KDD dataset and outperforms traditional methods in efficiency and detection performance [17]. This paper introduces a hyperdimensional computing-based framework for IoT network anomaly detection on the NSL-KDD dataset, achieving 91.55% accuracy and outperforming traditional methods in high-dimensional, complex environments [18]. This paper explores the feasibility of detecting cognitive load in complex virtual reality training environments by using eye-tracking data and machine learning–based analysis methods to interpret user responses [19].

This paper proposes a hierarchical Type-II fuzzy controller optimized via the imperialist competitive algorithm for load frequency control, enhancing robustness and dynamic response in power systems [20]. This paper develops an adaptive-intelligent quadcopter controller based on Brain Emotional Learning, demonstrating improved performance over classical controllers in both adaptive and non-adaptive modes [21]. This paper presents a cellular teaching-learning-based optimization approach that effectively addresses dynamic multi-objective problems, demonstrating superior results compared to state-of-the-art algorithms in evolving environments [22].

Despite notable progress, current DOA control methods continue to face critical limitations in delivering robust performance across diverse patient populations. Many strategies, including neural network and fuzzy controllers, lack systematic optimization tailored to varied physiological profiles, resulting in inconsistent BIS regulation. While FOPID controllers offer improved control capabilities, their performance heavily depends on the precise tuning of fractional orders and gains—a process that becomes computationally intensive without the aid of advanced AI-based optimization. Furthermore, the application of WOA-optimized FOFPID controllers in DOA remains relatively underexplored, particularly concerning their robustness in managing physiological uncertainties across multiple patient models. These gaps underscore the necessity for a robust, AI-enhanced control strategy that effectively integrates fuzzy logic, fractional-order control, and WOA-based optimization to deliver reliable and precise BIS regulation. The proposed study aims to fill this gap by developing and evaluating a WOA-optimized FOFPID controller across diverse patient models, as elaborated in the forthcoming sections on Theoretical Framework & Proposed Methodology and Simulation Results, to push the boundaries of automated anesthesia delivery.

### III. SYSTEM DESCRIPTION

Dynamic of the anesthetic drug (Propofol) is described by Pharmacodynamic (PD) and Pharmacokinetic (PK) model. PD describes how to distribute the drug in the body and PK describes the effect of the drug, considering its concentration in the body. In this model, the drug is distributed in three parts of the body, the first part of the Oregon that involves a lot of blood vessels such as the brain and heart, the second part of the muscle tissue, and the third part of the Orange, such as the skin and Bones. PD is modeled as following equations (1-3):

$$\dot{x}_1(t) = -(k_{10} + k_{12} + k_{13})(t) + k_{21}x_2(t) + k_{31}x_3(t) + u(t) \quad (1)$$
$$\dot{x}_2(t) = k_{12}x_1(t) - k_{21}x_2(t) \quad (2)$$
$$\dot{x}_3(t) = k_{13}x_1(t) - k_{31}x_3(t) \quad (3)$$

In equation (1-3), $k_{ij}$ (i, j = 1,2,3) is an anesthetic drug transfer parameter from i-th to j-th. The metabolic parameter ($k_{10}$) is the removal of the anesthetic drug from the part one. $x_i$ also indicates the amount of anesthetic drug in the i-th section. Infusion rate of the drug is U(t), $k_{ij}$ is determined by the physiological constants of each patient's body. The parameters are given in the following equations:

$$V_1 = 4.27 \, [L] \, , V_2 = 18.9 - 0.391(\text{age} - 53) \, [L]$$
$$, V_3 = 238 \, [L]$$

$$Cl_1 = 1.89 + 0.0456 \, (\text{Weight} - 77) + 0.0264 \, (\text{Height} - 177) - 0.0681 \, (\text{LBM} - 59)$$

$$Cl_2 = 1.29 - 0.024 \, (\text{age} - 53) \, , Cl_3 = 0.836$$

Where, the concentration of drug in each compartment is represented by $Cl_i$ and the volume of each compartment is represented by $V_i$, Lean Body Mass (LBM) and k_ij is measure is expressed as following:

$$\text{For male:} \quad \text{LBM} = 1.1 * W - 128 * (W^2/H^2)$$

$$\text{For female:} \quad \text{LBM} = 1.07 * W - 148 * (W^2/H^2)$$

$$k_{10} = Cl_1/V_1 \, , \, k_{10} = Cl_2/V_1 \, , \, k_{13} = Cl_3/V_1 \, , \, k_{31} = Cl_2/V_2 \, , \, k_{31} = Cl_3/V_3$$

PD is shown in the following equations (4-5):

$$\dot{c}_e(t) = k_{e0}\left(C_p(t) - C_e(t)\right) \quad (4) \, , C_p = x_1/V_1$$

$$BIS(t) = BIS_0 \left(1 - \frac{C_e^\gamma(t)}{C_e^\gamma(t) + EC_{50}^\gamma(t)}\right) \quad (5)$$

Where the effect site compartment concentration is represented by $C_e$, $k_{e0}$ is the constant of $C_e$, $EC_{50}$ is the concentration of drug in which half of the drug's greatest effect is observed in the patient, $\gamma$ is the nonlinear term the equation (Hill coefficient) and $BIS(t)$ is the variable to show DOA during anesthesia and goal of the system is to maintain its value in 50.

### IV. CONTROLLER DESCRIPTION

*A. Fractional order fuzzy PID controller*

Conventional PID controller is the most used controller in the industry, which has three coefficients for determining the input signal to the system, as given in formula 6:.

$$U_{PID}(s) = \left(K_P + \frac{K_I}{S} + K_D S\right).e(s) \quad (6)$$

Where, $U_{PID}(s)$ is the control signal, $K_P$ is the proportional gain, $K_I$ is the integral gain, $K_D$ is the derivative gain of conventional PID controller. In FOPID controller, derivative and integral calculations are performed in fractional order. In this paper, Fractional Order Modeling and Control Toolbox for Matlab (FOMCON) is used to calculate the fractional order derivative and integral. FOPID formula is expressed in (7):

$$U_{FOPID}(s) = \left(K_P + \frac{K_I}{S^\alpha} + K_D S^\beta\right) \cdot e(s) \quad (7)$$

Where, $U_{FOPID}(s)$ is the control signal, $S^\beta$ is the fractional-order derivative and $\frac{1}{S^\alpha}$ is the fractional-order integral, According to formula (7), FOPID has two degrees of freedom more than the PID controller, so it can perform better.

In FOFPID controller, the PID controller coefficients, which were constant, can change according to the system conditions to improve the performance of the control system. Also, the existence of a fuzzy inference system, due to its nature, can control systems with uncertainty and nonlinearity. In this study, the proposed fuzzy inference system has two inputs (error and error derivatives) and three outputs (controller coefficients), each of which has five membership functions. Figure 1 shows the structure of FOFPID controller.

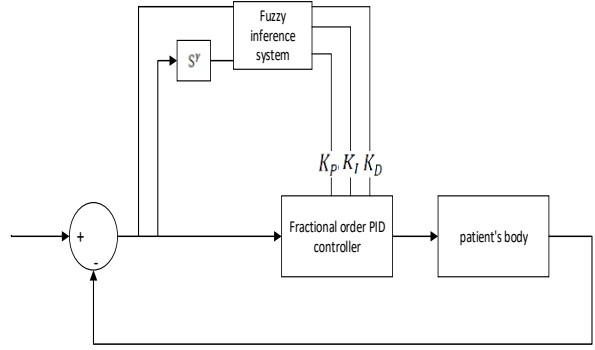

Fig. 1. Structure of FOFPID controller

B. *Whale Optimization Algorithm (WOA)*

The WOA algorithm is a completely new meta-heuristic algorithm that emulates humpback whale hunting behavior. The algorithms are the simulation of hunting behavior randomly with the best search agent for chasing prey and using a spiral motion to simulate the mechanism of the bubble-net feeding. This algorithm starts with a set of random answers. In each step, the search agents update their position according to the choice of the search agent. The operation of WOA will be described in the following formulas.

Searching prey phase is presented in equations (8-9):

$$D = |C * X_{rand} - X| \quad (8)$$

$$X(t+1) = X^*(t) - A * D \quad (9)$$

Where A, C are variable coefficient vectors, $X$ is the position vector (position of the whale), $X_{rand}$ is the random position vector (position of random the whale), vector $X^*$ is the best position vector (search agent). A, C are also obtained according to the following formula (10-11):

$$A = 2 * a * r - a \quad (10)$$

$$C = 2 * r \quad (11)$$

Where, $a$ is the parameter that initial value is 2 and decreases linearly to zero and $r$ is the random number [0-1].
Encircling prey phase is expressed in equations (12-13):

$$D = |C * X^*(t) - X(t)| \quad (12)$$

$$X(t+1) = X^*(t) - A * D \quad (13)$$

And the mathematical model of bubble-net attacking is given in the following formula (14):

$$X(t+1) = |X^*(t) - X(t)| * e^{bl} * \cos(2\pi l) + X^*(t) \quad (14)$$

Where $b$ defines as the shape of logarithmic spiral and l is the random number in [-1 , 1]. Bubble-net attacking is used when p (random number) is less than 0.5, and if p is larger than 0.5 Encircling prey phase is used [13]. If A⩾1 , then searching prey is used (exploration phase), otherwise Encircling prey and bubble-net attacking (exploitation phase) is used.

In this study, membership functions, rules and scaling factors of the fuzzy controller and orders of integral and derivative controllers (α,β,γ) is optimized and obtained via WOA. Appropriate cost function of the control system is given in equations. (15-17):

$$ITAE = \int_0^t t.|e(t)|\ dt \quad (15)$$

$$IAE = \int_0^t |e(t)|\ dt \quad (16)$$

$$Cost\ function = IAE_1 + ITAE_1 \quad (17)$$

## V. SIMULATION AND RESULT

In this section, Performance of FOPID and FOFPID controller have been investigated. Physiological logical model of 8 patients has been shown in table 1 and figures of optimized fuzzy membership functions are shown in figure 2.

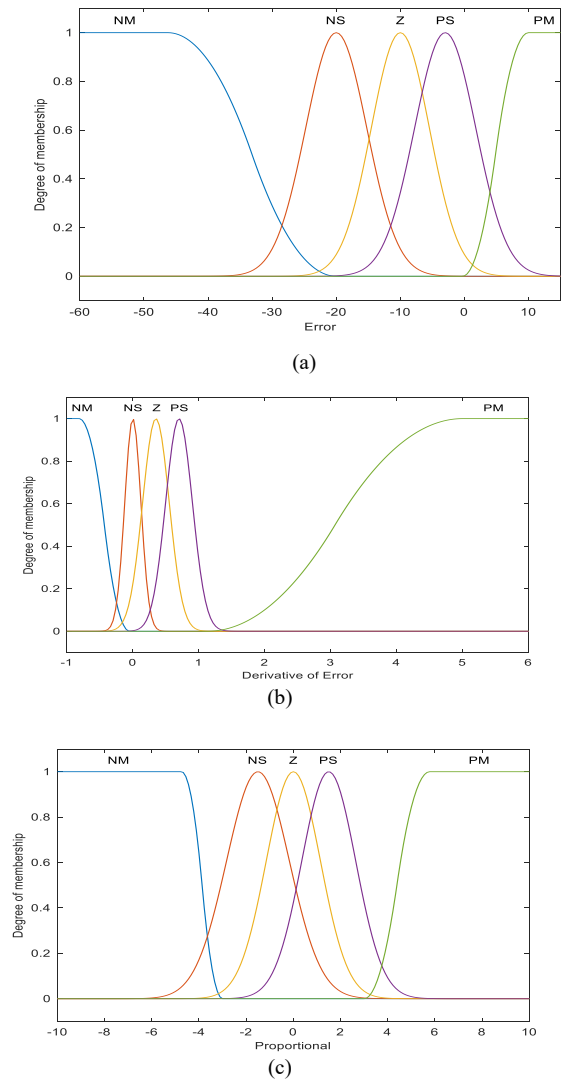

Fig. 2. First input type-I fuzzy sets (a), second input type-I fuzzy sets (b), output fuzzy sets (c) after optimization via WOA

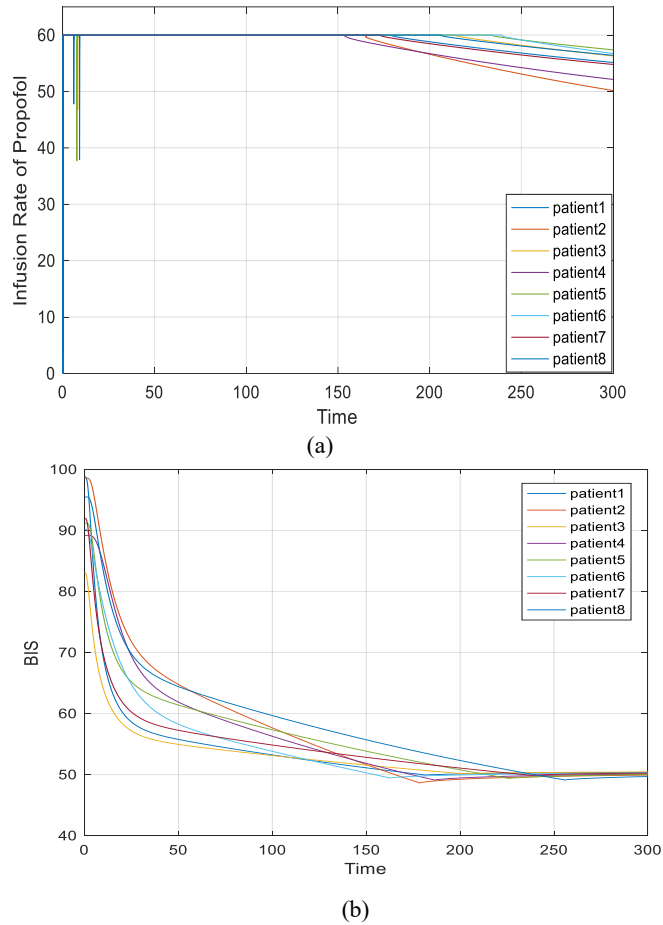

Fig. 3. BIS of 8 patient using FOPID (a) and FOFPID (b)

The results of using the two proposed controllers are shown in figure 3. According to figure 3 and table 2, Patients generally reach in desired value in about 3 minutes to start surgery and illustrated that Both controllers can control the system, But the FOFPID in comparison FOPID shows better behavior in the transition state and steady state. To describe precisely, settling time and steady-state error is far less. So FOFPID has a much better response than the FOPID. The infusion rate of Propofol (signal controller) is shown in figures 4.

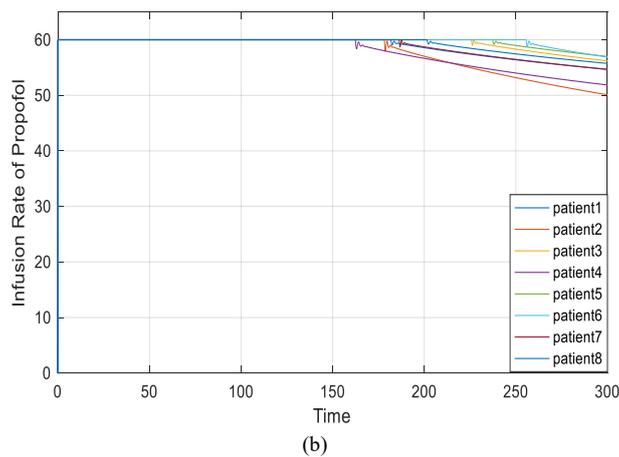

(b)

Fig. 4. The propofol infusion rate of 8 patient using FOPID (a) and FOFPID (b)

Figure 4 indicates that the controller uses all the control signals (infusion rate of Propofol) allowed to reach the desired value earlier, and when BIS reaches the desired level, it decreases).

VI. CONCLUSION

This study represents FOPID, FOFPID controller as closed-loop delivery of Propofol which is optimized by WOA to control DOA. Regarding the simulations performed and the results obtained, both controllers can control the nonlinearity and uncertainty of the system. The FOFPID controller has a better performance than the FOPID controller in transient and steady-state behavior due to its more degree of flexibility because of the more tunable parameters. For future work, the more number of patients can be used to accurately and better design of the controller.